\documentclass[]{aa}
\usepackage{natbib} 
\usepackage{graphicx}
\usepackage{txfonts}

\newcommand{\FBDY}{{\tt A$^\mathtt{sf}$}}
\newcommand{\FBDZ}{{\tt B$^\mathtt{nosf}$}}

\newcommand{\sdrop}{$\sigma$-drop}
\newcommand{\sdrops}{$\sigma$-drops}

\newcommand{\rcor}{$R_\mathrm{CR}$}
\newcommand{\rhilr}{$R_\mathrm{hILR}$}
\newcommand{\rvilr}{$R_\mathrm{vILR}$}

\newcommand{\msol}{M$_{\sun}$}
\newcommand{\sfrunit}{M$_{\sun}\ \mbox{yr}^{-1}$}

\begin{document}

\title{Formation of young boxy/peanut bulges in ringed barred galaxies}
\titlerunning{Formation of young boxy/peanut bulges}

\author{Herv\'e Wozniak\inst{1} \and L\'eo Michel-Dansac\inst{2,3}}

\offprints{H. Wozniak, \email{herve.wozniak@obs.univ-lyon1.fr}}

\institute{Universit\'e de Lyon, Lyon, F-69000, France ;
Universit\'e Lyon 1, Villeurbanne, F-69622, France ;
 Centre de Recherche Astrophysique de Lyon, 
Observatoire de Lyon, 9 avenue Charles Andr\'e, Saint-Genis Laval cedex, F-69561, 
France ;
 CNRS, UMR 5574 ; Ecole Normale Sup\'erieure de Lyon, Lyon, France
\and
IATE, CONICET, OAC, Universidad Nacional de C\'{o}rdoba,
 Laprida 854, X5000BGR, C\'{o}rdoba, Argentina 
 \and
 Consejo Nacional de Investigaciones Cient\'{i}ficas y T\'{e}cnicas,
 Argentina 
}
\date{Received / Accepted}

\abstract 
{} 
{We investigate whether the formation mechanism of boxy and
  peanut-shaped (B/PS) bulges could depend on the gas content of the galaxy.}
{We have performed N-body simulations with and without
  a gaseous component. In the second case star formation/feedback
  recipes have also been implemented to create new stellar
  populations.}
{As in many previous studies, in our N-body collisionless
  simulation, the B/PS is due to the classical break in the $z$ mirror
  symmetry lasting roughly 200~Myr.  When a gaseous component and star
  formation recipes are added to the simulation, the bulge-growing
  mechanism is quite different. The young
  stellar population that is born in the thin gaseous disc rapidly
  populates vertical resonant orbits triggered by the combined effects
  of the linear horizontal and vertical ILRs. This leads to a B/PS bulge
  mainly made of stellar material younger than the surrounding
  population. 

  The non-linear analysis of the orbital structure shows that the main
  orbit family responsible for the B/PS is not the same in the two
  cases. The 2:2:1 orbits prevail in the collisionless simulation
  whereas additional asymmetrical families contribute to the
  B/PS if a dissipative component is present and can form new
  stars. We found that 2:3:1 and 2:5:1 orbits trap a significant
  fraction of the mass.

  A flat ringed discy stellar component also appears simultaneously
  with the thickening of the young population. It is due to the star
  formation in a nuclear gaseous disc located in the central kpc,
  inside the ILR, and accumulated there by the torques exerted by the
  large-scale bar. Remarkably, it remains flat throughout the
  simulation although it develops a nuclear bar, leading to a
  double-barred galaxy.  }
{We predict that two populations of B/PS bulges could exist and even
  coexist in the same galaxy.
}

\maketitle

\section{Introduction}

How galactic bulges form is one of the leading questions for the
galaxy formation theories. A consensus is begining to emerge
\citep[e.g.][]{a05} that two scenarios of bulge formation compete. The
older scenario is the formation by an initial gravitational
collapse or, in a more recent scenario by a series of minor
mergers in a similar way to massive elliptical galaxies.  The second
scenario relies on the secular evolution of stellar discs.  This led
\citet{kk04} to make a distinction between ``pseudobulges'' that are
``bulges'' formed through secular evolution, and ``classical'' bulges,
those with round smooth isophotes that show no discy structure in the
central regions and thus are built up through mergers or a dissipative
collapse.

This dichotomy has been slightly refined by \citet{a05} who proposed
to further split the pseudobulge category into two classes.  The first
one contains boxy-peanut shaped bulge (B/PS bulges hereafter) that are
due to the vertical orbital structure of stellar bars seen edge-on
(\citealt*{cs81,p84,cdfp90,pf91}).  The frequency of B/PS bulges is
high: 45\%\ of all bulges are B/PS, while amongst those the exact
shape of the bulge depends mainly on the viewing angle to the bar
\citep{ldp00a,ldp00b}.  The observed incidence of B/PS bulges is
however consistent with that expected if they are associated with
relatively strong bars. Near infrared observations \citep{baadbf06}
have also revealed several B/PS features in `classical' bulges.  As
shown by N$-$body simulations, true peanuts are bars seen side-on,
i.e.\ with the major-axis of the bar roughly perpendicular to the
line-of-sight. For less favourable viewing angles, the bulge/bar looks
boxy, and if the bar is seen end-on it looks almost
spherical. Stronger bars also lead to more prominent peanut shapes, as
demonstrated observationally \citep[e.g.][]{ldp00b} and theoretically
\cite[e.g.][]{ba05}.

According to \citet{a05}, ''disk-like'' bulges (DL bulges hereafter)
belong to a second class. They are formed by star formation occurring
in the gaseous inflow possibly driven by a stellar bar. Bulges formed
according to this scenario can have observational properties
attributed normally to stellar discs (exponential photometric
profiles, blue color, substructures like spiral arms, nuclear bars,
circumnuclear rings, etc). In general, they can contain a measurable
amount of gas, as well as a young stellar population sometimes
distributed in bright spots. According to their mode of formation, DL
bulges should have a much smaller scaleheight than B/PS or classical
bulges since the gas distribution is rather flat even in the central
galactic region. It is thus very questionable to name these structures
``bulge'' \citep[see discussion by][]{a05}.

>From the simulation point of view, it has been considered that DL
bulges deserved little attention. However, they did appear under
several other names (exponential bulges, nuclear discs, double-bars,
bar-within-bar, etc) in a number of numerical simulations or
theoretical works \citep[i.e.][]{sb89, sbf90, fm93, fwrmb96, wcef03,
  wc06}. The main difficulty in self-consistently studying the
formation mechanism of such bulges or other related central/nuclear
morphological features comes from the fact that numerical codes have
to include a dissipative component and recipes to mimic star formation
and feedback processes, even using very simple rules and/or crude
approximations. Apart from some peculiar cases, collisionless N$-$body
codes are thus not able to reproduce DL bulges.

DL bulges could be associated with central velocity dispersion drops
\citep{wcef03} such as those observed by \citet{egcflpw01} and
\citet{Metal03}. The \sdrops\ result from the concentration of new
stars toward the centre, and because this population of new stars is
newly formed from the low-dispersion gas component, its velocity
dispersion is much lower than for the old population. This effect is
also amplified by the fact that the gas dispersion also drops toward
the centre (and therefore the new stellar component too). This is due
to the strong accumulation of gas toward the centre, in a nuclear
disk, where dissipation is stronger than elsewhere, and where the gas
therefore cools down efficiently. The stellar velocity dispersion
could remain low even if the star formation rate is rather low
\citep{wc06}, e.g. 1~\sfrunit, that is the same order of magnitude as
the typical gaseous mass inflow rate into nuclear rings \citep{rvt97}.
This is additional evidence of a small DL bulge scaleheight.

We decided to investigate the edge-on properties of B/PS and DL bulges. As
stated above, this kind of study must perform N$-$body simulations
including gas and star formation recipes. Typical simulations are
presented in Sect.~\ref{sec:model}. In the rest of this paper we
mainly discuss the formation mechanisms of both the B/PS and DL bulges
and their dynamical properties. In particular, we show that the young
stellar population is able to form a B/PS bulge via a slightly
different dynamical mechanism than classical B/PS bulge
(Sect.~\ref{ssec:thickening}). The formation of this young B/PS bulge
is accompanied by a flat nuclear disc
(Sect.~\ref{ssec:flatdisc}). This could be interpreted through a linear
and non-linear dynamical analysis of the families of resonant orbits
(Sect.~\ref{ssec:resonances}).

It should be stressed that young populations are much brighter for a
few 10$^7$ yr than old populations, so a DL bulge may also result from
the light contrast, and not only from the mass distribution.  In
accompanying papers we will therefore address the issue of detecting
young B/PS bulges predicted by our simulations \citep{mw08} and their
kinematics (in preparation). For this purpose, we will use
photometrically calibrated simulations including the absorption by the
dust distribution in the disc.


\section{Description of the numerical simulations}
\label{sec:model}

\begin{figure*}
\centering
\resizebox{\hsize}{!}{\includegraphics{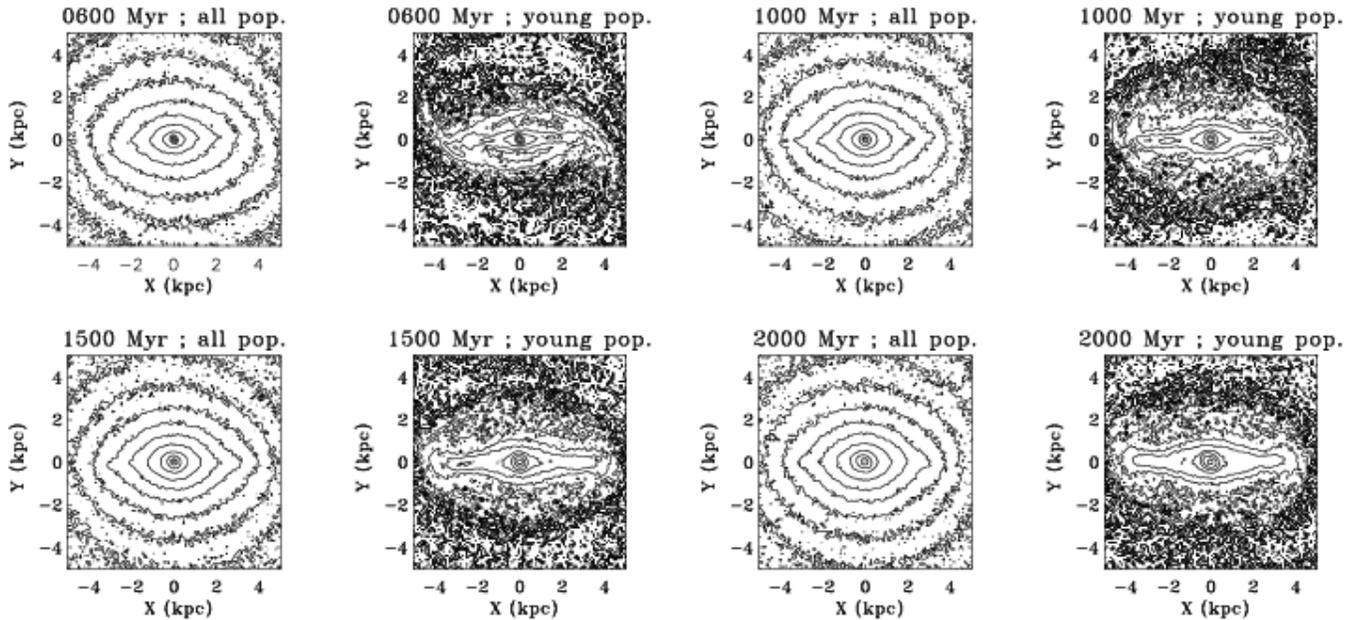}}
\caption{Face-on projected mass distribution in the central 10~kpc for
  $t=600$, 1000, 1500 and 2000~Myr. The snapshots have been rotated to
  align the bar with the $x$-axis. 'young pop.' means only the
  population of particles created during the run.}
\label{fig:contour}
\end{figure*}

\begin{figure*}
\centering
\resizebox{\hsize}{!}{\includegraphics{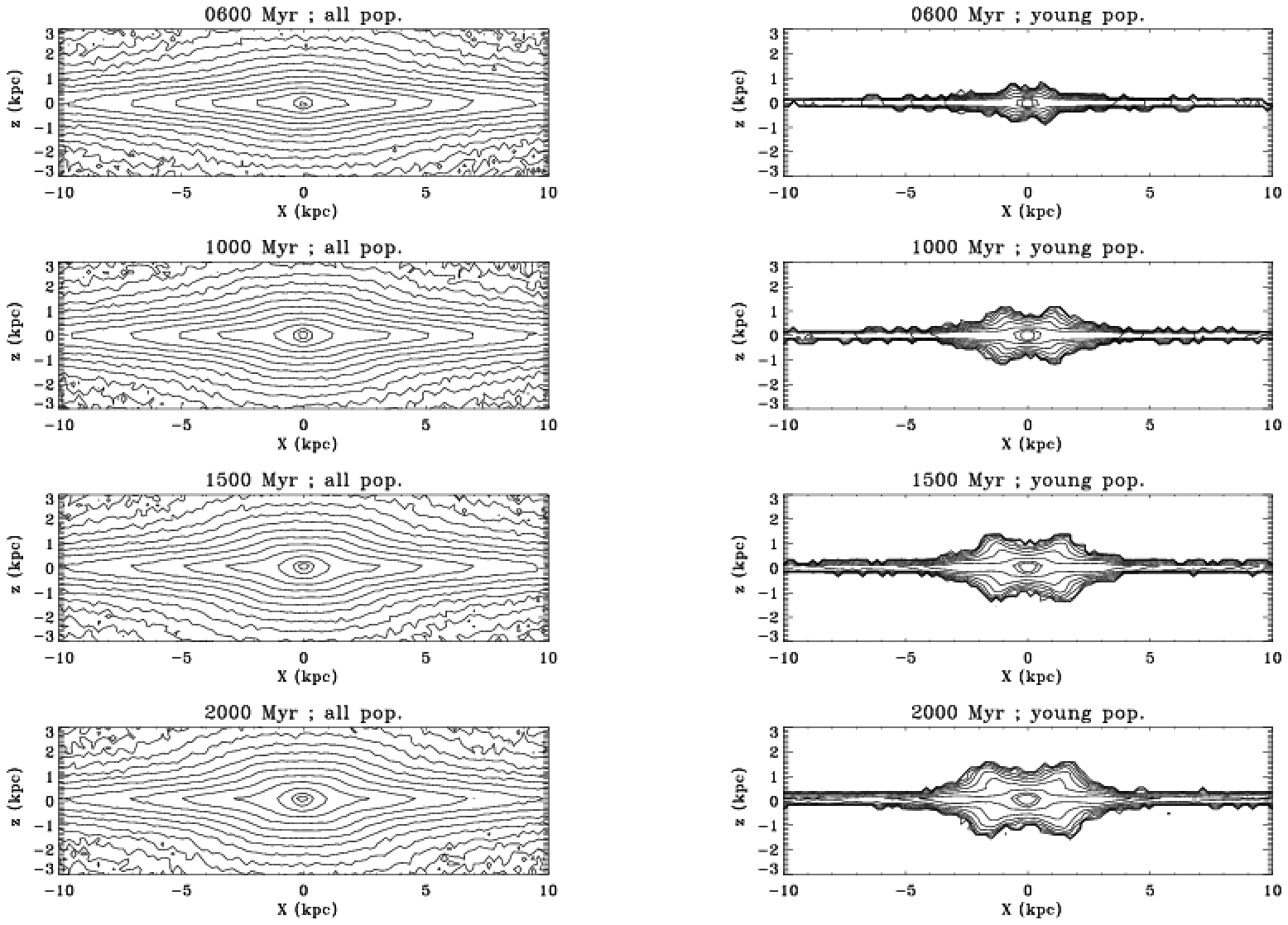}}
\caption{Left panel: edge-on projected mass distribution in the
  central 20$\times$6~kpc for $t=600$, 1000, 1500 and 2000~Myr. The
  snapshots have been rotated to align the bar with the
  $x$-axis. Right panel: only the population of particles created
  during the run.}
\label{fig:contoureo}
\end{figure*}

For clarity, we will concentrate on a single case extracted from a
dozen such simulations of various resolutions and initial setups. The
generic simulation, named \FBDY\ hereafter, is thus representative of
our database. Other examples can be found in \citet{mw08}.

An {\em initial} stellar population is set up to reproduce a typical
disc galaxy. Positions and velocities for $2.5\,10^6$ particles are
drawn from a superposition of two axisymmetrical \citet{mn75} discs of
mass $M_1$ and $M_2$ (cf. Table~\ref{tab:simul}), of scalelengths $1$
and $3.5$~kpc and a common scaleheight of $0.5$~kpc. Initial velocity
dispersions are computed solving numerically the Jeans equations
according to the \citet{H93} method. The method was extended to take
account of the presence of a dissipative component when solving for
the stellar equation.  The initial velocity dispersion was chosen to
be anisotropic with $\sigma_r = \sigma_z$ and $\sigma_{\theta}^2 =
\sigma_r^2 \kappa^2 / (4\Omega^2)$, where $\sigma_r$,
$\sigma_{\theta}$ and $\sigma_z$ are three components of the velocity
dispersion along respectively the radial, azimutal and vertical
directions and $\kappa$ and $\Omega$ are respectively the radial and
angular epicyclic frequencies. The resulting initial Q parameter in
the central 500 pc radius increases quickly with radius from 1 to 1.5
and then slowly increases up to 2.3 in the external part of the disc.
 
The initial disc radius is 30~kpc.
The gaseous component of run \FBDY\ is represented by 50\,000
particles for a total mass of $1.1\,10^{10}$~M$_{\sun}$ distributed in
a 6~kpc scalelength Miyamoto-Nagai disc.

\begin{table}
\centering
\caption{List of runs. Masses are in $10^{11}$\msol\ units.}
\label{tab:simul}
\begin{tabular}{@{}llllllll@{}}
\hline 
Model &  SF & $M_1$ & $M_2$ & $M_g$ \cr
      &     &       &       &       \cr
\hline
\FBDY&  on  & 0.1   & 1.0   & 0.11\cr
\FBDZ&  off & 0.11  & 1.1  &  0.  \cr
\hline
\end{tabular}
\end{table}

A reference run of pure collisionless particles, named \FBDZ, has been
computed to carry out various dynamical comparisons. For this
homologous run, $M_1$ and $M_2$ have been proportionally scaled so as
to keep the same total mass and spatial distribution as \FBDY.

The evolution is computed with a particle--mesh N-body code, derived
from the original version of the Geneva group \citep{pf93,fb93}, which
includes stars, gas and recipes to simulate star formation. The broad
outline of the code is the following: the gravitational forces are
computed with a particle--mesh method using a 3D log--polar grid with
$(N_R, N_\phi, N_Z)=(60,64,312)$ active cells. The smallest radial
cell in the central region is 36~pc large and the vertical sampling is
50~pc. The extent of the mesh is 100~kpc in radius and $\pm 7.8$~kpc
in height.  The hydrodynamics equations are solved using the SPH
technique.  Since we used a polar grid and we need an accurate
determination of the forces in the central region, we have improved
the pre-computation of self-forces by subdividing each cell in $(n_r,
n_\phi, n_z)=(32,6,6)$ subcells. Self-forces are then linearly
interpolated before being subtracted from the gravitational
forces. The spatial resolution and force accuracy are thus much higher
than in any of our previous studies based on the same code
\citep{fwrmb96,wcef03,mw04,Hetal05,mw06,efwfms06,wc06}.

The star formation process is based on Toomre's criterion for the
radial instability of gaseous discs (cf. \citealt{mw04} for more
details). When star formation is active, the radiative cooling of the
gas is computed assuming a solar metallicity. In
Figs.~\ref{fig:contour} and \ref{fig:contoureo} we display the face-on
and edge-on views of the bar region of \FBDY\ for four times which
will be used throughout this paper. They have been chosen as being
illustrative of the bulge evolution. At the end of the simulation
($t\approx 3000$~Myr), the total number of particles is roughly
$3.2\,10^{6}$ for the stellar component and $30\,000$ for the gaseous
one. 45\%\ of the gas has been transformed into stellar particles,
mainly in the central 10~kpc.

The main effect of a live dark halo (except to flatten the rotation
curve of the disc at a large distance) is to permit the exchange of
angular momentum with the stellar disc. The rate and the amplitude of
these exchanges depend on the velocity dispersion of both the disc and
the halo, and on the relative halo mass
\citep[e.g.][]{DS00,a03,VK03}. The stellar disc could lose between a
few \%\ and 40\%\ of its angular momentum mainly through
resonances. Depending on the rate at which the stellar disc losses its
angular momentum, the bar grows quite differently. Considering
\citet{msh06} simulations as representative, roughly 2/3 of the
angular momentum loses by the bar-unstable part of the stellar disc is
absorbed by the halo, the rest going to the outer disc. Most of these
exchanges happen during the buckling of the bar. Afterward, the halo
absorbs all the angular momentum lost by the disc, leading to a second
phase of buckling \citep{msh06}. The lack of a live dark halo in our
simulations thus has the main consequence that we are not able to find
any second buckling phase.

However, our simulations are not completely devoid of vertical
exchanges since the Miyamoto-Nagai density distribution allows us to
build substantially inflated bulges, which is the case with our choice
of parameters that leads to an S0-like initial stellar
distribution. Thus the evolution of the particles confined near the
$z=0$ plane is also driven by vertical exchange of angular momentum.
Moreover, this paper is a first report of the effect of a gaseous
component with active star formation on the vertical structure of
bars. It is important to be able to disentangle the effect of the
dissipative component from that of the halo on the disc
evolution. Thus, simulations with a live dark halo, that are still in
progress, will be reported elsewhere.

%
\section{Central region thickening}
\label{ssec:thickening}
In the case of pure N$-$body simulations, i.e.\ made of collisionless
particles only, whenever a disc galaxy forms a bar, a B/PS bulge
develops in a few dynamical times. This process was firstly studied
by \citet{cs81}, confirmed by \citet{cdfp90} and \citet{rsjk91}, and
recently reanalyzed by many authors
\citep[][etc]{mwhdb95,am02,a02,a03,a05,msh06}.  A stellar bar starts
thin and then buckles out of the plane after a short phase of mirror
symmetry breaking in the $z$ direction \citep{pf91}. Thus, initially
asymmetrical with respect to the equatorial plane, the bar finally
tends towards symmetry \citep{msh06}. Only a fraction of the whole bar
buckles \citep{a05}, so that the outermost part of the bar stays thin
and planar.

\begin{figure}
\resizebox{\hsize}{!}{\includegraphics{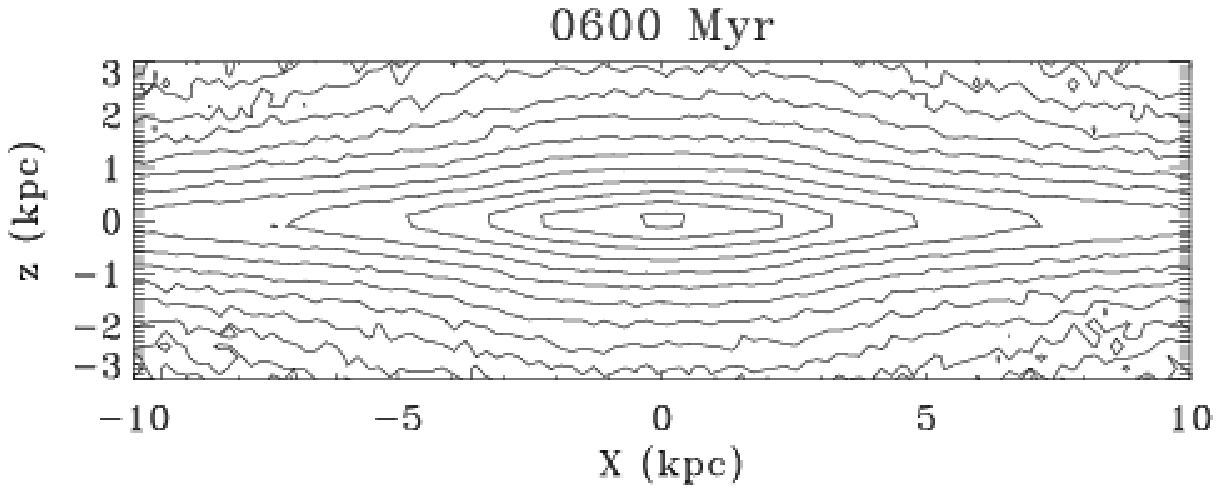}}
\resizebox{\hsize}{!}{\includegraphics{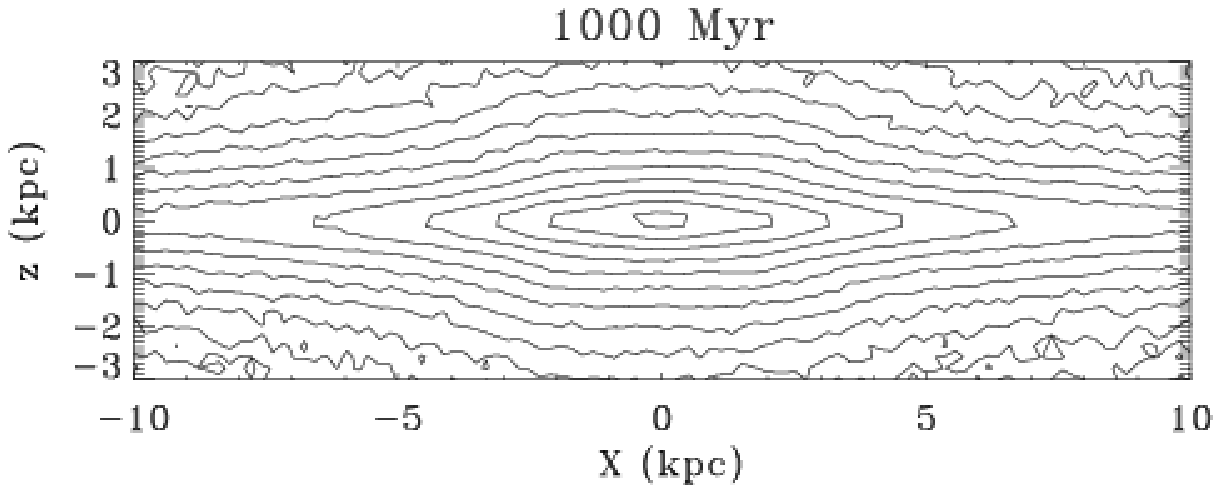}}
\resizebox{\hsize}{!}{\includegraphics{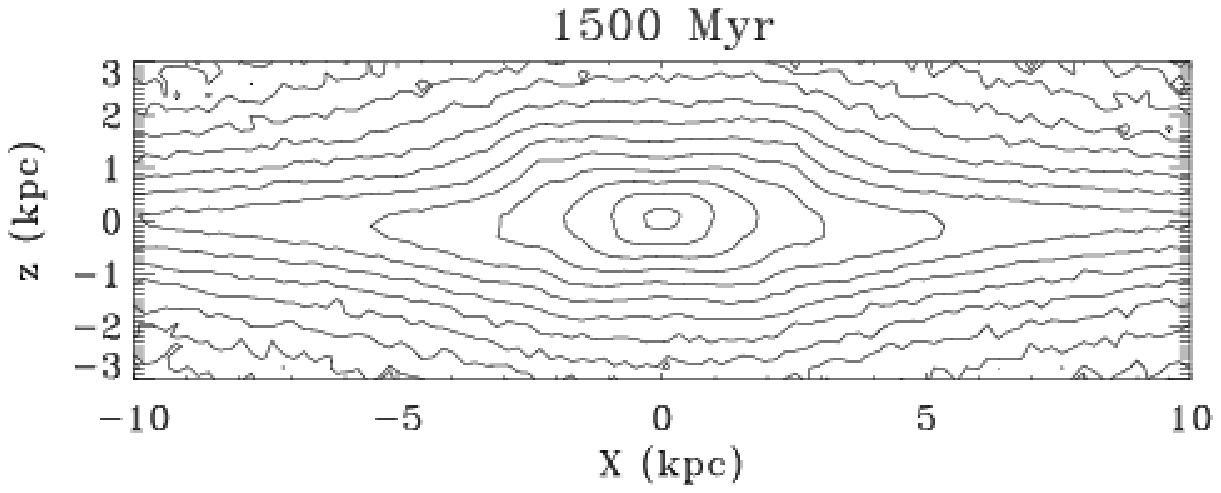}}
\resizebox{\hsize}{!}{\includegraphics{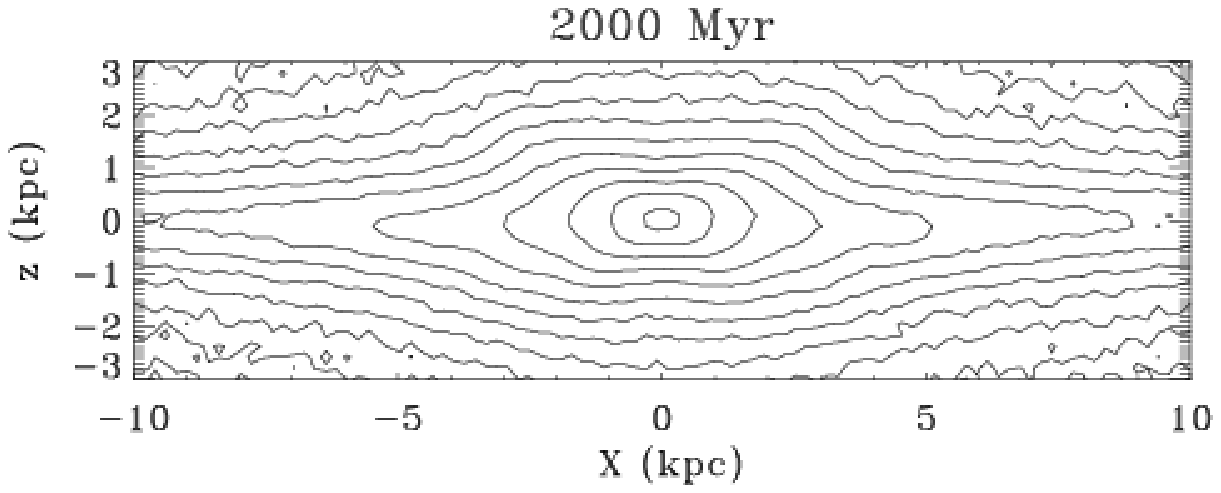}}
\caption{Edge-on projected mass distribution of run \FBDZ\ in the
  central 20$\times$6~kpc for $t=600$, 1000, 1500 and 2000~Myr.}
\label{fig:contoureoref}
\end{figure}

This is once more what happens for the bar in \FBDZ\ (see
Fig.~\ref{fig:contoureoref}). Between $t\approx 1400$ and 1600~Myr the
vertical thickening starts due to a break in the $z$ mirror
symmetry. It is only for $t>1600$~Myr that the $z$-distribution
becomes more symmetrical.

In the case of \FBDY\, most of the young stellar population lies in a
razor-like central disc during the first 450~Myr. This is due to
the small vertical scaleheight of the gas distribution that remains
thin because of its dissipative nature. It is well-known that such a
stellar razor-thin disc is highly unstable \citep[e.g.][]{ms94}.  Indeed,
450~Myr after the beginning of the young disc formation, the most
central part of the disc starts to thicken out of the equatorial
plane. In roughly a bar rotation period, the vertical distribution
becomes {\em symmetrically} peanut shaped over the central 2~kpc
(Fig.~\ref{fig:contoureo}, right panel, at $t=600$~Myr), while the
young bar is approximately 8~kpc long
(Fig.~\ref{fig:contour}). However, at this time, the total mass of the
central disc still being low, the thickening process has no detectable
effect on the global mass distribution (Fig.~\ref{fig:contoureo}, left
panel).

\begin{figure}
\resizebox{\hsize}{!}{\includegraphics{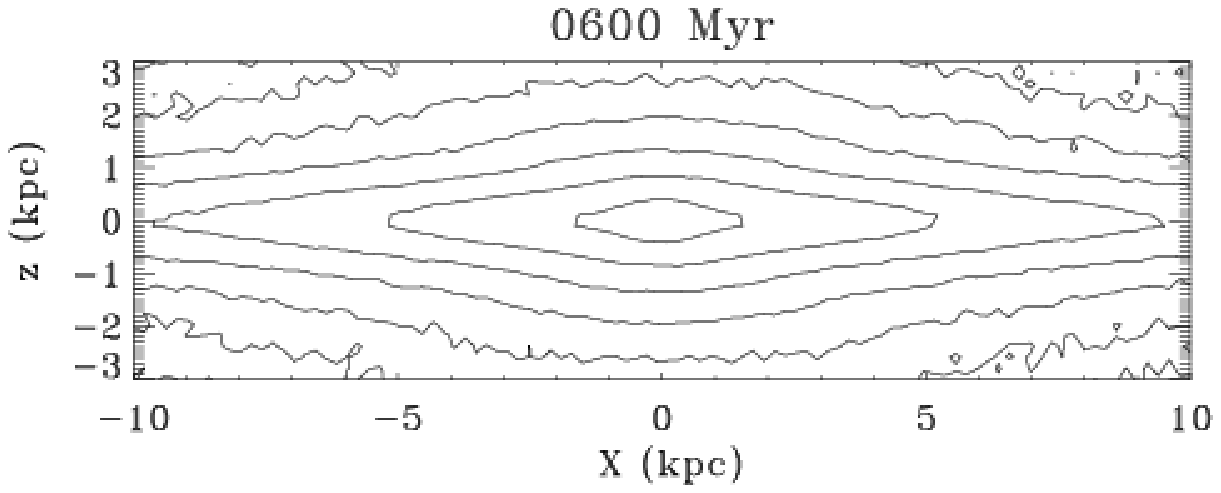}}
\resizebox{\hsize}{!}{\includegraphics{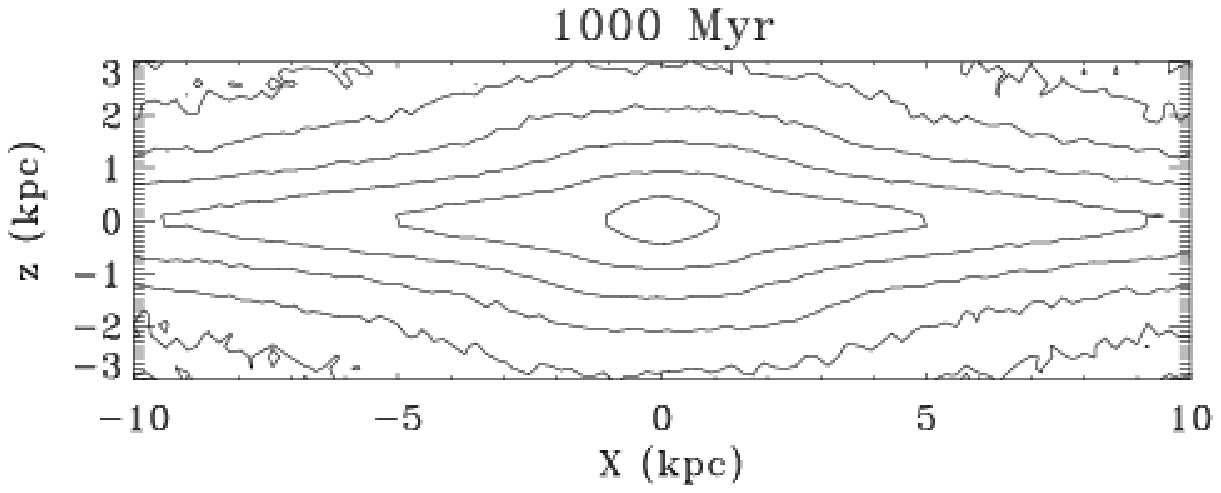}}
\resizebox{\hsize}{!}{\includegraphics{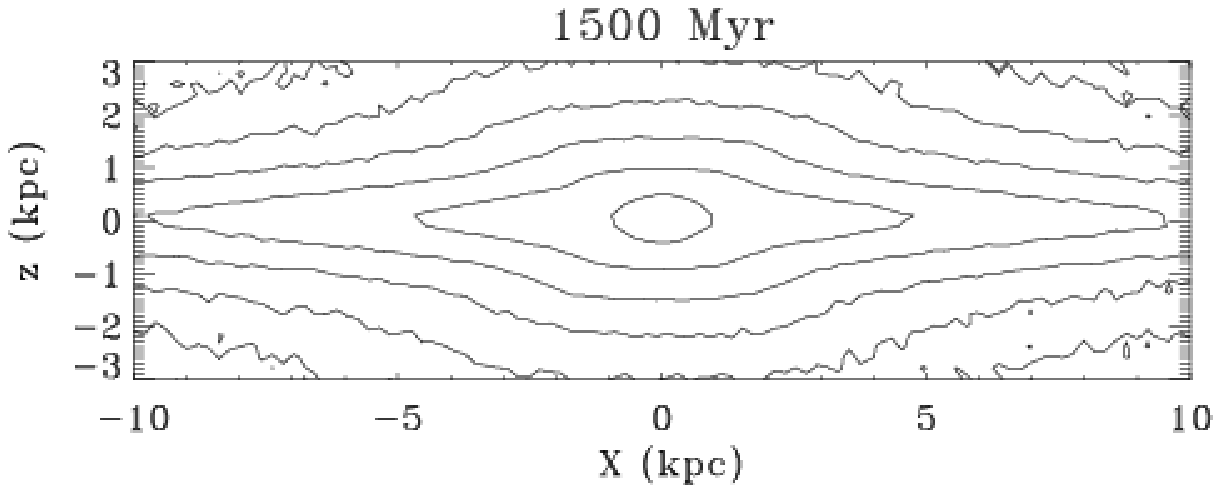}}
\resizebox{\hsize}{!}{\includegraphics{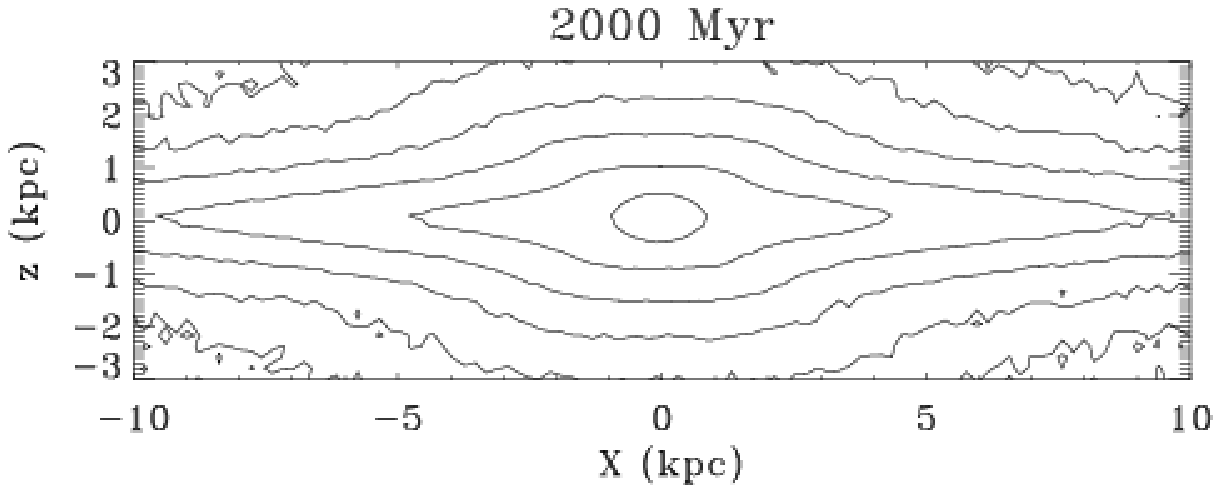}}
\caption{Edge-on projected mass distribution of the {\it initial}
  population in the central 20$\times$6~kpc for $t=$600, 1000, 1500
  and 2000~Myr.}
\label{fig:contoureoold}
\end{figure}

Afterward, the peanut-shape widens out as the young disc continuously
evolves and increases in mass. The thick part of the disc doubles its
radial size in less than 1~Gyr. The vertical scaleheight also
increases with time leading to a well-developed peanut-shaped bulge
for $t > 1500$~Myr. It is noteworthy that the total mass distribution,
hence including both the initial and young population, plainly
displays a B/PS bulge. This is mainly due to the mass of the young
population that amounts to a significant fraction of the whole mass at
that time. Indeed, the B/PS bulge is clearly much less marked if we
only display the mass distribution of the initial population
(Fig.~\ref{fig:contoureoold}).

Disentangling the two causes that could be responsible for the
increase in boxiness of the {\em initial} population is not so
obvious. Indeed, as the mass of the young population trapped in the
B/PS increases, the gravitational potential deviates more and more
from its initial quasi-spherical shape. This leads to a vertical mass
redistribution of the initial population since particles can be
trapped by orbits associated with vertical resonant families.  But the
initial disc population could be also unstable towards vertical
instabilities \citep{rsjk91}. This is supported by looking at the
homologous simulations \FBDZ\ that also develop a B/PS bulge
(Fig.~\ref{fig:contoureoref}). For \FBDZ\ we said above that the
growth mechanism is the same as for all other collisionless
simulations of cold stellar discs performed so far.

However, a few differences between \FBDZ\ and \FBDY\ should be
noticed:
\begin{enumerate}
\item{At the end of the buckling phase, the extent of the boxy region
  is slightly wider for \FBDZ\ than \FBDY. Moreover, \FBDZ\ is much
  more peanut-shaped than \FBDY, especially at $t=2000$~Myr and
  beyond. In Sect.~\ref{ssec:resonances} we attempt to explain this in
  terms of resonances.}
\item{The buckling is clearly asymmetrical for \FBDZ\ between $t=1400$
  and 1600 Myr, whereas it remains symmetrical for \FBDY. If any
  asymmetry in the mass distribution appears during the evolution, its
  scaleheight should be less than the vertical resolution of the code
  (i.e.\ 50~pc). Indeed, since the $z$-distribution of the
  \FBDY\ young population remains always symmetrical with respect to
  the equatorial plane, the potential well created by the young
  population is also permanently symmetrical. It thus could be more
  difficult to break the $z$-symmetry for \FBDY\ than for \FBDZ.  This
  difference supports the interpretation that vertical resonant orbits
  are populated.  }
\end{enumerate}

In principle, we cannot exclude the possibility that the thickening of
the young population might be asymmetrical because low-order bending
modes, those that can give an asymmetrical-shaped mass distribution,
can be erased by our 50~pc vertical resolution. Indeed, it has been
shown by \citet{ms94} that finite grid effects, especially the
vertical mesh resolution, could stabilize the low-order bending modes
in the case of a perfectly thin disc. However, the young disc created
in all our simulations is always embedded in the initial stellar
population which has a much higher scaleheight (initially 500~pc,
increasing with time). The vertical distribution of the initial
stellar population also remains symmetrical with respect to the equatorial
plane (cf. Figs.~\ref{fig:contoureoref} and
\ref{fig:contoureoold}). Moreover our vertical resolution is much
higher than for the Merritt \& Sellwood simulations. The vertical
oscillating frequencies of particles thus are not dominated by the
mesh spacing as is the case for \citet{ms94} simulations.

\section{Flat \emph{nuclear} disc}
\label{ssec:flatdisc}
\begin{figure}
\resizebox{\hsize}{!}{\includegraphics{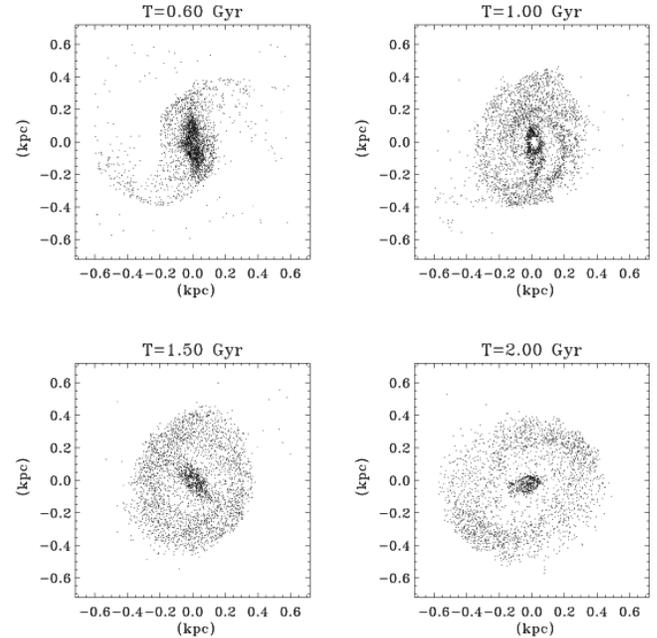}}
\caption{Face-on gas mass distribution in the central region where the
  flat \emph{nuclear} stellar disc made of the young population
  develops for \FBDY. The gas particle distribution has not been
  convolved by the SPH kernel to emphasize the ring structure.}
\label{fig:nucleardisc}
\end{figure}

As discussed above, a thin stellar disc cannot remain flat for a long
time, such as the \FBDY\ initial young central disc.  However, for
\FBDY\, even if a significant part of the young stellar disc thickens
rapidly, there is a small {\em nuclear} component which remains flat
for a very long time (Fig.~\ref{fig:nucleardisc}), i.e.\ until the end
of the simulation ($\approx 3.5$~Gyr). This nuclear component is made up
of a young stellar population formed in the gaseous disc accumulated
in that region because of the torques exerted by the large-scale
stellar bar. The formation of this flat \emph{nuclear} structure is
thus concomitant with the thickening of the \emph{central} part
discussed in Sect.~\ref{ssec:thickening}. After 2 Gyr, the stellar
mass amounts to roughly $7.25\times 10^{9}$ \msol\ in a cylinder of $R = 0.5$
kpc and $z \pm 0.5$ kpc. The young population accounts for roughly
34\% of the total mass.

The nuclear disc quickly develops a small bar with its own pattern
speed, almost 10 times higher than the large-scale bar. The nuclear
bar is encircled by a circumnuclear ring
(Fig.~\ref{fig:nucleardisc}). The nuclear bar appears at $t\approx
575$~Myr and is long-lived, although at the resolution of our
simulations (between 36~pc at the centre and 100~pc at 1~kpc), it
periodically dissolves in a spiral-like structure. The detailed study
of the morphological, dynamical and kinematical properties of the
nuclear bar and the circumnuclear ring deserves a dedicated paper and
thus will be publish elsewhere \citep{w08}. Hereafter we ignore the
internal structure of the nuclear disc.

The existence of flattened and rapidly rotating {\em nuclear} stellar
discs has been predicted by \citet{sb89} and further studied by
\citet{sbf90}. They showed that such stellar discs could remain
flattened for a long time since the two-body relaxation is a slow
process.  \citet{cb04} detected such a nuclear disc in roughly 1/3 of
their sample of 24 edge-on B/PS galaxies. They have given clear
evidence that these discs exhibit a \sdrop\ in their centre.  They
also speculated on the possibility that such discs are formed through
gas inflow and subsequent star formation.  The whole initial young
central disc discussed in Sect.~\ref{ssec:thickening} cannot be
analogous to the flat disc of \citet{sb89}, but its nuclear component
obviously has similar properties.

>From the kinematical point of view, the nuclear component is undoubtedly
associated with the \sdrop\ phenomenon.  \citet{wcef03} showed
that such a young nuclear stellar disc is responsible for the \sdrop\
since it is the kinematical signature of stars that have been born from a
dynamically cold gaseous component. The presence of the nuclear bar
and transient spiral structure in \FBDY\ marginally increases the
radial velocity dispersion, but the effect on the line-of-sight
velocity dispersion remains weak so that a \sdrop\ should be visible
in double-barred galaxies. \FBDY\ thus confirms the potential
relationship between nuclear disc and \sdrop\ in B/PS bulges as
suggested by \citet{cb04}.

The respective sizes of the nuclear disc and peanut-shaped bulge are
given in Table~\ref{tab:sizes}. The two components enlarge as the galaxy
evolves but the ratio of the box over the disc sizes also slightly
increases with time. The two phenomena, whose the common initial
cause is the presence of a young stellar population formed in the
inflowing gaseous material driven by a large-scale bar, seem
dynamically distinct. Both morphological structures seem to evolve
independently from each other. However, the main driver of the
internal dynamics is well-known to be the large-scale bar and the set
of resonances associated with its rotation pattern.

\begin{table}
\centering
\caption{Rough sizes of the box (from rotation axis to corner) and the
  central stellar disc (radius in kpc) determined on the mass
  distribution of the young population alone. The radius of main linear
  resonances for \FBDY\ and \FBDZ. hILR is the horizontal inner
  Lindblad resonance, vILR the vertical one and hUHR is the horizontal
  ultra harmonic resonance.}
\label{tab:sizes}
\begin{tabular}{@{}llllll@{}}
\hline 
      & 600 & 1000 & 1500 & 2000 \cr
      & Myr & Myr  & Myr  & Myr  \cr
\hline
Box and disc sizes &     &       &       &    \cr
disc  & 0.27 & 0.43 & 0.50  & 0.50 \cr
boxy  & 0.67 & 1.17 & 1.40  & 1.59 \cr
boxy/disc & 2.48 & 2.69 & 2.80 & 3.18 \cr
\hline
Resonance radii of \FBDY &     &       &       &    \cr
hILR  & 1.27 & 1.61 & 1.89  & 1.80 \cr
vILR  & N/A  & 1.34 & 1.68  & 1.81 \cr
hUHR  & 3.57 & 3.82 & 4.21  & 4.00 \cr
\hline
Resonance radii of \FBDZ &      &      &       &      \cr
hILR  & 1.61 & 2.24 & 2.45  & 2.84 \cr
vILR  & N/A  & N/A  & N/A   & 2.95 \cr
hUHR  & 4.69 & 5.12 & 5.25  & 5.26 \cr
\hline
\end{tabular}
\end{table}

\section{Horizontal and vertical resonances}
\label{ssec:resonances}
\subsection{Linear analysis}
\label{ssec:linear}

Numerous authors have tried to correlate the size of morphological
structures to the dynamical resonance locations. Circumnuclear and
outer rings seem to be correlated with the location of, respectively,
the inner Lindblad resonance (ILR) and outer Lindblad resonance (OLR)
(cf. \citealt{BC96}). The ratio of nuclear bar length to that of the
large-scale bar could be similar to the ILR to corotation (CR) ratio,
the nuclear bar corotation being dynamically coupled to the
large-scale bar ILR (e.g. \citealt{RS99}).  However, some other
simulations did not show such coupling (e.g. \citealt{hse01}); this
matter is still under debate.

To determined the location of the \emph{linear} CR and ILR dynamic
resonances, we computed the circular orbit frequency $\Omega$ and the
radial epicyclic frequency $\kappa$ as \citep{P90}:
\begin{equation}
\Omega^2(r) = \left\langle
\frac{1}{r}\frac{\partial\Phi}{\partial r}
\right\rangle
\end{equation}
\begin{equation}
\kappa^2(r) = \left\langle
\frac{\partial^2\Phi}{\partial x^2}+
\frac{\partial^2\Phi}{\partial y^2}+
2\,\left( \frac{1}{r}\frac{\partial\Phi}{\partial r} \right)
\right\rangle
\end{equation}
where $\Phi$ is the gravitational potential and $\langle\cdots\rangle$
stands for an azimuthal average. 

Strictly speaking, these frequencies predict the oscillation
frequencies of the orbits in the axisymmetrical case only. They do not
provide any indication of whether families of periodic orbits do follow
such oscillations when the bar growth breaks the axisymmetry. However,
a number of previous orbital studies \citep[cf.][and discussion
  therein]{mw06, wm07} suggest that the epicyclic
approximation could lead to an acceptable estimation of the resonance
locations displayed in Table~\ref{tab:sizes}, in particular
if we are mainly interested in their evolution rather than their
accurate absolute position. For instance, using a careful integration
of orbits to accurately compute $\Omega$ and $\kappa$, \citet{mw06}
found that the error on \rcor\ remains within 10\%. This technique has
been used to study a few snaphots when looking for higher order
resonances (cf. Sect. \ref{ssec:nonlinear}).

\begin{figure}
\resizebox{\hsize}{!}{\includegraphics{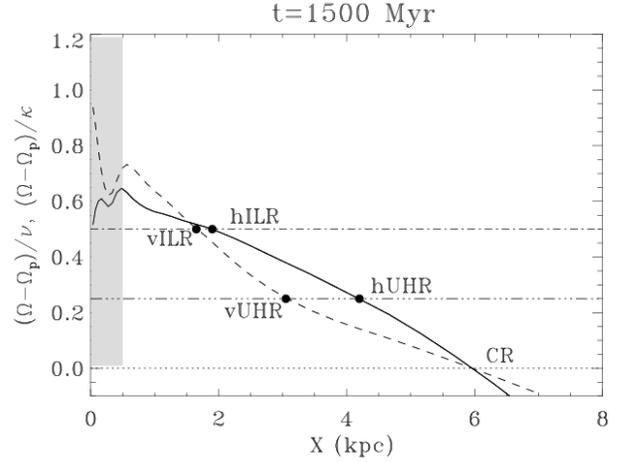}}
\caption{Resonance diagram $(\Omega-\Omega_\mathrm{p})/\kappa$
(horizontal resonances, full line) and
$(\Omega-\Omega_\mathrm{p})/\nu$ (vertical resonances, dashed line) at
$t=1500$~Myr. Location of ILR (dot-dash), UHR (triple dot-dash) and CR
(dot) are plotted. The shaded box depicts the region occupied by the
nuclear disc.}
\label{fig:resonances}
\end{figure}

The flat nuclear disc is entirely inside the linear ILR of the large-scale
bar, its radius being roughly 0.5~kpc at $t=1500$~Myr. Indeed,
\rhilr\, the horizontal ILR (hILR) radius computed using the linear
approximation is $\approx 1.9$~kpc from the centre
(cf. Fig.~\ref{fig:resonances}).  The gaseous component which also
forms a nuclear disc occupies the same region as the young stellar
nuclear disc.  During the evolution of the disc, it is well known that
the large-scale bar slows down, leading \rhilr\ to increase with
time. Hence one could imagine that the size of the nuclear disc
increases proportionally to \rhilr\ but in fact its growth saturates
after roughly 1.5~Gyr. The nuclear disc does not entirely fill the region
encircled by the hILR. Indeed, the radius where $\Omega-\kappa/2$ is
maximum is expected to be the limiting dynamical radius. In
Fig.~\ref{fig:resonances}, the limiting radius is roughly where
$(\Omega-\Omega_p)/\kappa$ is maximum, that is quite close to an
$\Omega-\kappa/2$ local maximum.

The combined effects of the horizontal and vertical ILRs cause the
planar orbits lying in the equatorial plane to be destabilized
\citep{cdfp90}, but the B/PS region extends outside the
ILRs. However, as already outlined by \citet{pf91}, in order to
observe a well defined boundary in the B/PS it is necessary that
families of orbits supporting the shape cease to exist or, at least,
become unstable beyond some well-defined height. \citet{pf91}
suggested by default that corotation could be the limiting
resonance. However, at $t=1500$~Myr, the vertical ILR (vILR) is
located at $\approx$1.7~kpc, which is very similar to the distance of
the corner of the B/PS from the axis of rotation
(cf. Table~\ref{tab:sizes}) that could be the extent of orbit
families associated with the vILR. This obviously suggests that the
ILRs, and their associated families of resonant orbits, are
responsible for limiting the extent of orbits in the vertical
direction. Since it has been shown that the horizontal UHR delineates
the bulk of the bar in the equatorial plane \citep{mw06, wm07}, the
combined effects of UHR and ILRs play a very important role in shaping
the central region of disc galaxies.

\begin{figure}
\resizebox{\hsize}{!}{\includegraphics{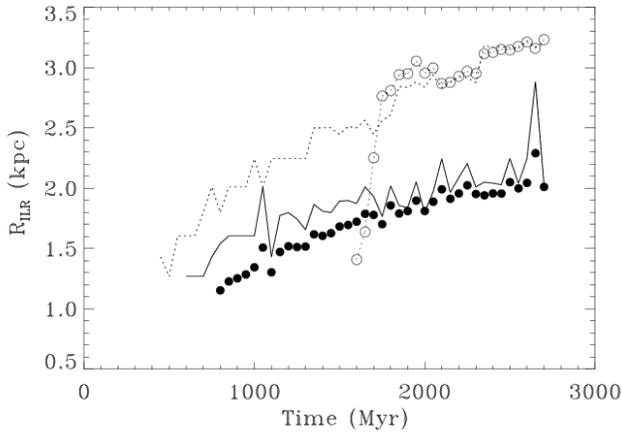}}
\caption{Evolution of horizontal (lines) and vertical (dots) linear
  ILR radius for run \FBDY\ (black thick line and full symbols) and
  \FBDZ\ (dotted lines and opened symbols).}
\label{fig:evolresonances}
\end{figure}

At $t=1500$~Myr (Fig.~\ref{fig:resonances}) the
location of the hILR is very close to the location of the vILR
(\rvilr$\approx 1.7$~kpc at $t=1500$~Myr). This coincidence has been
observed by previous authors \citep{cdfp90, pf91} and it has
been suggested that this situation could have some consequences on the
local dynamics. As shown in Fig.~\ref{fig:evolresonances}, for \FBDY\
the coincidence occurs during all the run, \rvilr\ being always
slightly smaller than \rhilr, although the difference decreases with
time. The rapid changes in the nuclear mass distribution due to the
gas inflow imply rapid fluctuations of the ILR positions as well, the
fluctuation timescale typically being the local dynamical time (less
than 50~Myr).

For \FBDZ, the coincidence arises progressively
(cf. Fig.~\ref{fig:evolresonances}). Indeed, during the first 1.6~Gyr
there is no vILR. The vILR appears at a radius \rvilr$\approx 1.4$~kpc
and then increases to coincide with the hILR at $t\approx
2.1$~Gyr. We have said above that the B/PS bulge formation
for \FBDZ\ follows a different path than for \FBDY. Indeed, before the
vILR appears, the stellar disc buckles out {\em asymmetrically}. It is
likely only when families of orbits associated with the vILR appear
that the mass distribution could be symmetrized with respect to the
equatorial plane. This is also the case for \FBDY\ although both ILRs
appear simultaneously. The thickening of \FBDY\ is thus symmetric
because ILR resonant families exist from the beginning of the bulge
formation.

\subsection{Non-linear analysis}
\label{ssec:nonlinear}

To be able to discuss the potential effect of higher order resonances,
we have computed the orbital frequencies $\Omega$, $\kappa$ and $\nu$
of a representative sample of particles for a few selected
snapshots. We have applied a variant of the technique of
\citet{a03}. We have frozen the potential at a given time in the
simulation and then computed orbits in the inertial potential to
determine the principal frequencies using the technique of
\citet{ca98}. We used almost 50\,000 particles as initial conditions
chosen at random in a limited domain of the phase-space. Indeed, being
mainly focused on the resonances inside the bar, we restrict our
computation to the particles that \emph{a priori} reside in the region
encircled by the corotation. We a posteriori checked that the ($E_J$,
$r$) space ($E_J$ being the Jacobi constant) and time-averaged
Lindblad diagram ($E_J$, ${\bar L_z}$) are well sampled by this
choice. To discuss our results we will used the notation of
\citet{sw93} for closed (periodic) orbits. In this notation,
$m$:$n$:$l$ implies $m$ radial oscillations in the $(x,y)$ plane and
$n$ vertical oscillations in $z$ as the orbit achieves $l$ rotations
about the center.

\begin{figure}
\includegraphics[width=\hsize]{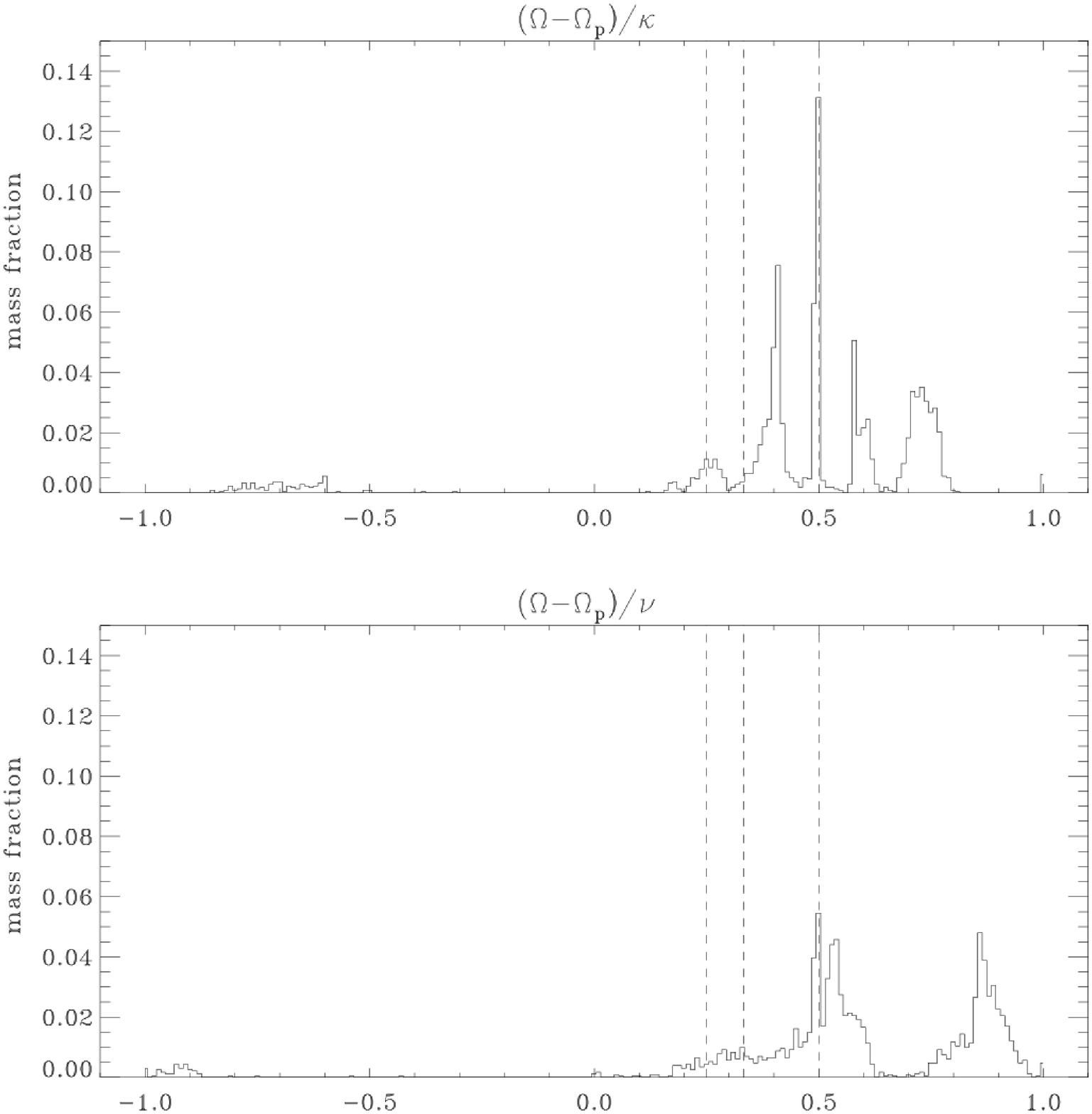}
\caption{Distribution of mass as a function of ${\cal
    K}=(\Omega-\Omega_p)/\kappa$ (upper panel) and ${\cal
    N}=(\Omega-\Omega_p)/\nu$ (lower panel) for \FBDZ\ at $t=2000$
  Myr. The total mass is normalised to unity. The 2:1:l (${\cal
    K}=0.5$), 3:1:l (${\cal K}=0.33$) and 4:1:l (${\cal K}=0.25$)
  resonances are plotted as dotted lines. Multi-periodic orbits
  ($l>1$) are superposed on $l=1$ orbits. }
\label{fig:omka2000}
\end{figure}

\begin{figure}
\includegraphics[width=\hsize]{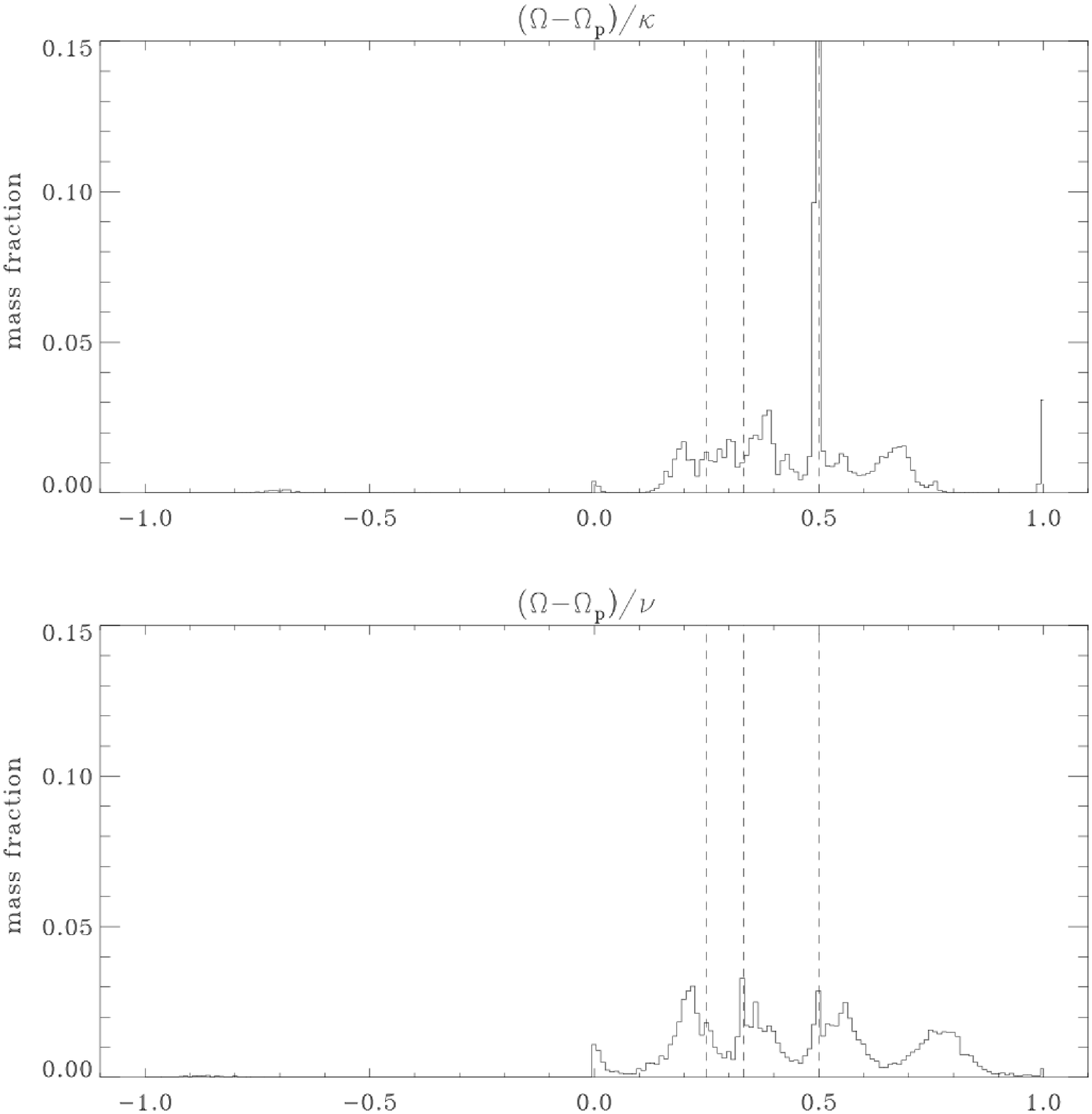}
\caption{As Fig. \ref{fig:omka2000} for \FBDY\ at $t=1500$. The peak
  at ${\cal K}=0.5$ has been truncated to 0.15 but extends to a value
  of 0.20.}
\label{fig:omka1500}
\end{figure}

We have selected the snapshot $t=2000$ Myr for \FBDZ\ which is the
moment when the B/PS shape is well established and the snapshot
$t=1500$ for \FBDY.  To easily identify periodic orbits that shape the
morphology of barred galaxies, we display the distribution of ${\cal
  K}=(\Omega-\Omega_p)/\kappa$ and ${\cal N}=(\Omega-\Omega_p)/\nu$ in
Figs.  \ref{fig:omka2000} to \ref{fig:omka1500}. Families of periodic
orbits thus have commensurable ratios ${\cal K} = l/m$ and ${\cal N} =
l/n$.

In these Figures, we have discarded orbits for which the maximal
absolute radial extent is less than 1 kpc to avoid contamination
by quasi-circular orbits that remain confined in the central nuclear
disc. These flat orbits do not participate in shaping the bulge and
mostly populate the regions ${\cal K}=1$ and ${\cal N}=0$. On the
contrary we did not try to filter out multi-periodic orbits ($l>1$)
so that they are mixed with $l=1$ orbits. 


For the 
two snapshots, the dominant family of orbits is related to the
equatorial 2:$n$:1 resonance, that is the external hILR (${\cal K} =
1/2$) since we have discarded as much as possible resonant families
associated with the innermost hILR.  However, both the $\cal K$ and
the $\cal N$ diagrams are quite different for \FBDY\ and \FBDZ\ and
deserve to be separately discussed, then compared.

For \FBDZ\ (Fig. \ref{fig:omka2000}), the dominant family is made of
2:2:l orbits (${\cal K}=1/2$ and ${\cal N}=1/2$).  These 2:2:1 orbits
are known to be responsible for the classical B/PS bulges
\citep{cdfp90,pf91} found in collisionless N-body simulations and 
have thus focused most attention under various names
(e.g. B$_{\mathrm z_1}$ for \citet{p84}, $z1$ for \citet{hpn93},
$x1v3$ for \citet{spa02}). These orbits dominate the $\cal K$ and
$\cal N$ distributions. However, since the mass fraction trapped
around the horizontal and the vertical frequencies are different, this
suggests that a fraction of the ${\cal K}=0.5$ peak is populated by
2:$n$:1 orbits with $n \neq 2$.

Several other non-negligible contributions are concentrated around
${\cal K}=0.25$ (4:$n$:1 orbits related to the UHR), $0.4,
\approx 0.6$ and $\approx 0.75$. It is much more difficult to uniquely
identify the orbit families responsible for these peaks but they mostly
contain multi-periodic trapped orbits. For instance we found 5:$n$:2
orbits that contribute to the ${\cal K}=0.4$ peak.

\begin{figure}
\begin{center}
\includegraphics[width=\hsize]{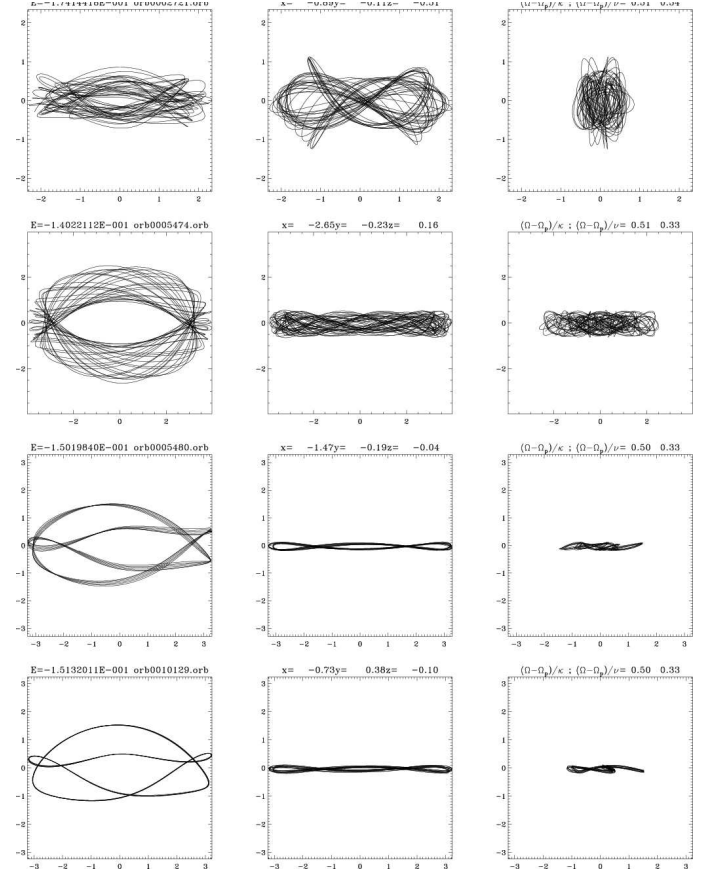}
\caption{Examples of dominant families of orbits for \FBDY\ at t=1500
  Myr. From top to bottom, a quasi 2:2:1 orbit and three different
  instances of 2:3:1 orbits.}
\label{fig:orb0}
\end{center}
\end{figure}

For \FBDY\ (Fig. \ref{fig:omka1500}), the distributions are very
different even if the main contribution is still concentrated around
${\cal K}=0.5$ as for \FBDZ. The fraction of mass (roughly 0.2)
trapped for ${\cal K}=0.5$ is however much higher than for
\FBDZ\ (0.13). The secondary peaks are less prominent than for
\FBDZ\ and they form a forest of frequencies betwen ${\cal K}=0.1$ and
$0.8$.

The vertical structure is also very different from the collisionless
run.  The mass trapped around ${\cal N}=1/3$ (the vertical 3:1
resonance) is slightly greater than for ${\cal N}=1/2$ resonant
orbits. ${\cal K}=1/2$ for these orbits, as for 2:2:1 orbits, so that
they must to be classified as 2:3:1 resonant orbits. A few examples
are shown in Fig. \ref{fig:orb0}.

The role of this family of orbits has been emphasized by \citet{hs96}
who demonstrated that it is connected to and amplified by \emph{the
  presence of a massive nuclear ring}. Indeed, \citet{hs96} showed
that the major periodic orbits within the corotation radius are
affected by the perturbation of a massive ring. The main family of
orbits that sustain a stellar bar (the so-called $x_1$ family in the
\citet{cp80} notation) becomes vertically unstable for several ranges
of Jacobi constant ($E_J$) values. In the $E_J$ domain where $x_1$
coexists with $x_2$ and $x_3$ families (that are respectively stable
and unstable elliptical-like orbits perpendicular to the bar and that
occupy the region between the two ILRs), two $x_1$ instability strips
provoke $z$ and $\dot{z}$ bifurcations to 2:2:1 families (the
so-called symmetrical BAN -- banana -- and asymmetrical ABAN --
anti-banana -- families) and 2:3:1 families.

\citet{hs96} showed that the width of these instability gaps depends on
the mass of the ring. As the ring mass is increased, the instability
strip responsible for 2:3:1 orbits grows in size and moves toward
lower $E_J$. In some extreme situations where the mass of the ring is
very high (e.g. $10^9$\msol) some instability strips appear for the
$x_2$ family.

2:3:1 orbits are symmetrical with respect to the $(x,z)$ plane and
antisymmetrical with respect to the $(y,z)$ one (or the converse)
unlike 2:2:1 orbits that are either symmetrical or antisymmetrical
about both the $(x,y)$ and $(y,z)$ planes. \citet{hs96} (their Fig. 7
and 8) and \citet{spa02} (their Fig. 9) display typical members of
this family.  It has been noted by \citet{spa02} and \citet{psa02}
that the 2:3:1 family (called $x1v3$ in these papers) as well as other
families associated with higher order vertical resonances (i.e. 2:4:1,
2:5:1 etc.), could play an important role in shaping small sized B/PS
bulges because the $z$ extent remains small whatever the values of
$E_J$ are. For \FBDY\ they appear to be a major contribution to the
shape of the young B/PS bulge whereas they are fully absent for \FBDZ.

\begin{figure}
\begin{center}
\includegraphics[width=\hsize]{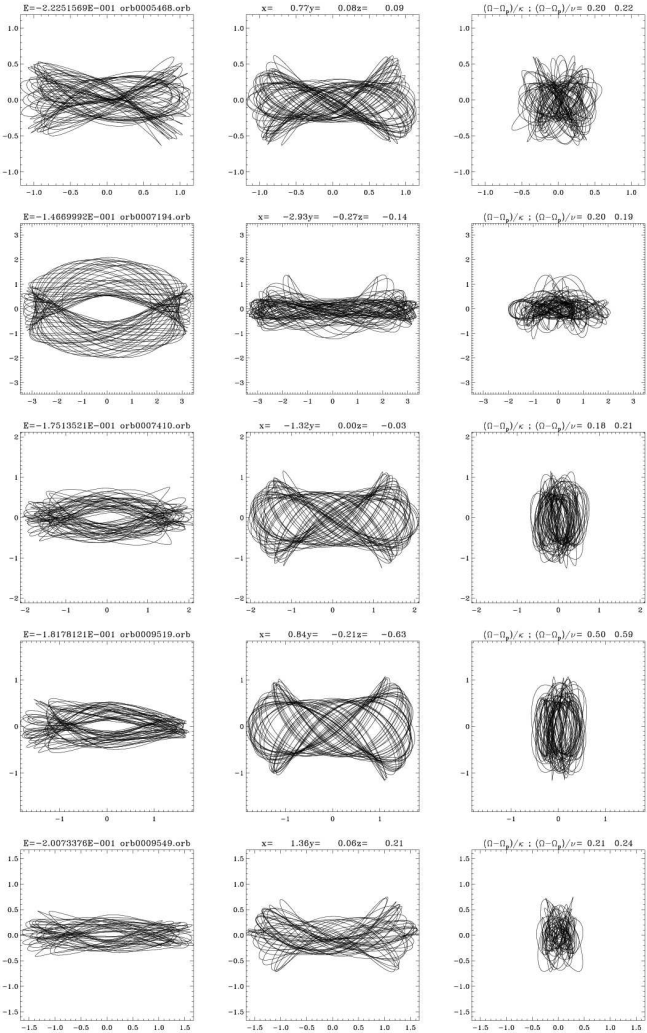}
\caption{Examples of other boxy-shaped families of orbits for \FBDY\ at t=1500
  Myr. }
\label{fig:orb1}
\end{center}
\end{figure}

For \FBDY\ the main family after 2:2:1 and 2:3:1 is the 2:5:1, visible
in Fig. \ref{fig:omka1500} around ${\cal N}=0.2$. As discussed by
\citet{psa02} this family is also a strong contributor to the B/PS
bulge. But, as for 2:3:1 orbits, they are absent in \FBDZ.  Many other
kinds of resonant orbits contribute to the B/PS. In Fig. \ref{fig:orb1}
we display a few examples. Their contribution is mainly concentrated
around ${\cal K}\approx 0.2$ and $0.66$ whereas ${\cal N} = 0.2$.

Even if the families of closed periodic orbits are the backbone of any
stellar bar, strictly speaking they occupy a null volume of the
phase-space. Only orbits trapped around the \emph{stable} families are
responsible for the shape of the bar. These trapped orbits have
fundamental frequencies slightly shifted from those of their parent
families so that they are responsible for the broadening of the
spectral lines around the commensurable value of $\cal K$ and $\cal
N$.

\section{Conclusions}
\label{sec:conclusions}
We have numerically investigated some dynamical properties of edge-on boxy or
peanut-shaped (B/PS) and disc-like (DL) (pseudo-)bulges. Our results
are summarised as follows:

\begin{enumerate}

\item{ We are able to confirm that the B/PS in N-body collisionless
  simulations is due to the classical break in the $z$ mirror
  symmetry.  However, in our numerical simulation that includes a
  gaseous component and star formation recipes, the bulge-growing
  mechanism is quite different from the pure N-body case. The young
  stellar population that is born in a thin gaseous disc rapidly
  populates vertical resonant orbits triggered by the combined effects
  of the horizontal and vertical ILRs. This leads to a B/PS bulge
  mainly made of stellar material younger than the surrounding
  population. The morphology and extent of \emph{young} B/PS bulges
  are significantly different from the classical B/PS bulge. We thus
  predict that two populations of B/PS bulges could exist and even
  coexist. They might be distinguished by deep photometric
  observations or careful stellar population analyses.}

\item{ In N-body collisionless simulations the main orbit family
  responsible for the B/PS is the 2:2:1. On the contrary, if a
  dissipative component is present and can form new stars, additional
  asymmetrical families contribute to the B/PS. In the case of our
  simulation, 2:3:1 and 2:5:1 orbits trap a significant fraction of
  the mass.  Their appearance could be linked to the massive
  circumnuclear ring.}

\item{ A \emph{flat} discy stellar component appears simultaneously
  with the thickening of the young population. It is due to star
  formation in the nuclear gaseous disc. Remarkably, it remains flat
  throughout the simulation (3.5~Gyr) although it develops a bar, as
  predicted by \citet{sb89}. This suggests that nuclear bars, embedded
  in large-scale counterparts, are flat structures. Both the stellar
  and gas discs are located well inside the ILR, limited by the radius
  where $(\Omega-\Omega_p)/\kappa$ is maximum.}

\end{enumerate}

\section{Acknowledgements}
We warmly thank Luis Aguilar for providing his code to compute orbital
frequencies and Daniel Pfenniger for fruitful discussions about the
computation of resonances.  Our computations were performed on the
CRAL 18 node cluster of PCs funded by the INSU ATIP \# 2JE014 and
several grants from the INSU Programme National Galaxie. LMD
acknowledges support from the Universidad Nacional Aut\'onoma de
M\'exico (UNAM) for part of this work, the University of Lyon, the
HORIZON project and the ECOS-Sud program \# A07U01, for financial
support for his visits during which this paper was submitted.

\end{document}